\documentclass[11pt]{article}
\usepackage{amsfonts,amscd,amsmath,cite,graphicx}

\def\mysection#1{\section{#1}\setcounter{equation}{0}}

\title{Quantized tension: \\
Stringy amplitudes with Regge poles and parton behavior}
\author{Oleg Andreev\thanks{andre@itp.ac.ru}
\\
L.D. Landau Institute for Theoretical Physics\\
Kosygina 2, 119334 Moscow, Russia\\ \\
Warren Siegel\thanks{siegel@insti.physics.sunysb.edu}\\
C.N. Yang Institute for Theoretical Physics\\
State University of New York, Stony Brook, NY 11794-3840, USA}
\date{}
\begin{document}
\maketitle
\begin{abstract}
We propose stringy hadronic amplitudes that combine some of the features of sister
trajectories and running tension.  By summing over string amplitudes
with varying Regge trajectories that have
integer tension and converging intercept, we obtain parton hard-scattering and
Regge soft-scattering behaviors, while preserving discrete poles in both
momentum and angular momentum.
\end{abstract}
\vspace{-13.5cm}
\begin{flushright}
hep-th/0410131    \\
LANDAU-04/HEP-A1 \\
YITP-SB-04-38
\end{flushright}
\vspace{12 cm}
%_______________________     I N T R O D U C T I O N_________________
\mysection{Introduction}

Hadronic physics can be divided into four regions of ``phase space": (1) low
energy, (2) spectrum, (3) high energy, small angle, and (4) high energy, large
angle. Low energy (including the low end of the spectrum) is described by many
methods, such as lattice Quantum ChromoDynamics, nonlinear sigma models, instantons, and
nonrelativistic quark models (which also handle mass differences within any
multiplet).  The parton model, and more accurately
perturbative QCD, describe high energy at large angles, and to a lesser extent
at small angles (total cross sections and related processes). Regge theory
describes the spectrum, and scattering for high energies at small angles, as
well as being consistent with low energy.

Regge theory directly relates the hadronic spectrum to the high-energy behavior of
scattering amplitudes at small angles.\footnote{For a review, see \cite{rev}.}
Amplitudes are described by a Regge trajectory $\alpha(t)$: The spectrum consists of
states of spin $J=\alpha(M^2)$ at mass $M$, while amplitudes go as
$s^{\alpha(t)}$. The requirement of a perturbation expansion whose lowest
order realizes this behavior only as poles in angular momentum implies \cite{ven-rev}
Dolen-Horn-Schmid duality \cite{dhs}, which is explicitly realized in string
theory. Experiment has verified duality qualitatively, and Regge high-energy
behavior up to $|\,t\,|$ of the order of $1\,\,\text{GeV}^2$, but the most striking
confirmation of Regge behavior is the appearance of the known hadrons on very
linear trajectories up to spins as high as $4$.

 From the beginning it was known that string amplitudes had exponential decay
at large transverse momenta, as seen in the fixed-angle limit of high-energy
scattering \cite{ven,sh}, and it was soon realized that this could not easily be
reconciled with the observed power-law behavior described by parton models
\cite{br}. One interpretation was that strings and QCD were
dual descriptions of the same physics, so that parton behavior in theories of
hadronic strings is nonperturbative, just as confinement is nonperturbative in
QCD. Thus at the very least an infinite summation of diagrams would be
required to obtain one description from the other. For example, in Regge
theory a cut produced from the exchange of multiple reggeons has a Regge slope
a fraction of that of the original reggeon \cite{rev}, so a summation can
produce an effective leading trajectory, running along the tops of the
trajectories of the pole and cuts, that has harder behavior in the appropriate
region \cite{sr}. A similar approach is to use the ``sister trajectory" poles
related to these cuts, also found in progressively higher-point amplitudes
\cite{htw}. Unfortunately such an approach is intractable, just as trying to
calculate the soft parts of perturbative QCD amplitudes by infinite
resummation of QCD graphs. Also, the hard limits obtained in both the cut and
sister approaches are not the usual power laws of the parton model.

An alternative method is to use the coordinate of a fifth dimension as an
effective running tension \cite{amp}. In fact, this can be realized along the
lines of the Anti-deSitter/Conformal Field Theory conjecture \cite{malda}.
The simplest models are built from old-fashioned Veneziano or Virasoro-Shapiro
amplitudes $A_n$ integrated over the tension with an appropriate weight factor
as \cite{ps, rest}\footnote{If one assumes that perturbation theory is a
topological expansion, then the full amplitude can be defined as
$\hat A_n(p_1,\dots ,p_n;\xi_i,\dots ,\xi_n)=\sum_{\chi=2}^{-\infty}\,
g^{-\chi} \hat A_n^{(\chi)}(p_1,\dots ,p_n;\xi_i,\dots ,\xi_n)$, where $\chi$
is the Euler number \cite{loops}.}
\begin{equation}\label{am}
\hat A_n(p_1,\dots ,p_n;\xi_i,\dots ,\xi_n)=
\int_{r_0}^\infty dr\,\,r^{3-\Delta_n}\,
A_n(p_1,\dots ,p_n;\xi_i,\dots ,\xi_n)\vert_{\alpha'\rightarrow\alpha'R^2/r^2}
\quad,
\end{equation}
where $p_i$'s and $\xi_i$'s are momenta and wavefunctions of
particles. $\Delta_n$, $R$ and $r_0$ are some parameters whose
meaning will be clarified shortly.\footnote{Note that one can think of
$\hat A_n$ as the Mellin transform of $A_n$, namely $\hat A_n=r_0^{4-\Delta_n}
\int_1^\infty
d\rho\,\rho^{-\omega-1}A_n\vert_{\alpha'\rightarrow\hat\alpha/\rho^2}$,
with $\omega=\Delta_n-4$ and $\hat\alpha=\alpha'R^2/r^2_0$. The factor
$r_0^{4-\Delta_n}$ simply adjusts the dimension.} We will call $A_n$ (without
any modification of $\alpha'$) a ``primary" amplitude to differentiate it from
$\hat A_n$.

Integrating from zero to infinity manifestly produces a scale invariant
amplitude, appropriate to some conformal field theory (for example, N=4 super
Yang-Mills), while putting a lower limit on the integration keeps that limit as
a unit of tension, breaking scale invariance, as appropriate to QCD. This
effectively produces a continuum of sister trajectories, but all for the
four-point amplitude (for example). The ``top" trajectory for positive
argument corresponds to the minimum-tension trajectory, while for negative
argument it corresponds to the infinite-tension (zero-slope) trajectory,
appropriate to a parton. Regge high-energy behavior comes from the smooth
joining region intermediate between these two top pieces.\footnote{For a
different approach to Regge physics also motivated by the AdS/CFT conjecture,
see \cite{pes} and references therein.}

Unfortunately, the ``smearing" of the trajectories replaces the particle poles
with cuts (as opposed to the usual distinct poles plus cuts required by unitarity):  Effectively this means that for any
spin the masses are continuous.  Similar behavior occurs in conformal
theories with nonvanishing mass:  A conformal transformation scales $p^2$, and thus the mass, so it is possible for
massive theories to be conformally invariant if the mass spectrum includes all positive real numbers.   In this case not all continuous masses
extend to zero once one has introduced the QCD tension as an integration limit, breaking the conformal invariance but leaving the continuous mass problem unresolved.  (Continuous mass is a possible problem in AdS/CFT, and related problems appear in membranes and subcritical closed strings.) In particular, this destroys the usual low-energy limit
(``pion physics").

In this paper we propose amplitudes that simultaneously have (1) a discrete
(integer-spaced) particle spectrum appearing on linear Regge
trajectories\footnote{In the
literature there exist models also providing discrete spectra (see \cite{sup} and
references therein). Their crucial differences from ours are: (1) the use of the supergravity
approximation, and (2) occurrence of continuous spectra as well as poles. Since their assumptions
do not seem fundamentally different from those of \cite{amp,ps,rest} (slight modification of the
metric to implement a cutoff in the fifth dimension), in general it is difficult to see how this
discrete spectrum can be made consistent with Regge behavior and allowed kinematic regions.}, (2) Regge behavior in the soft limit, and
(3) parton behavior in the hard limit. Continuous spectra are avoided by replacing the integral
over tension with a sum. The original spectrum is preserved by requiring the tensions to be
integer multiples of the original, as for sister trajectories. The correct
parton behavior follows from requiring that the trajectory intercepts (which
also have a type of integer constraint) converge. (For large integers, which
contribute to the parton behavior, the sum can be approximated as an
integral.) We do not provide a string Lagrangian for these amplitudes, but
propose them as a starting point, as was the case for the original string.

We propose these amplitudes to describe ``tree-level" behavior with respect to
both partons and hadrons, which seems the only way to perturbatively calculate
amplitudes that necessarily contain both hard and soft pieces. Generally, when
nonperturbative properties are important in a formulation of a theory, it is
an indication of the limitation of that formulation. For example, in the usual
formulation of Quantum ElectroDynamics, one first calculates classical (tree),
then perturbative (loop) contributions, and that is sufficient for observed
phenomena (except when corrections of a non-electrodynamical nature
contribute). On the other hand, in QCD nonperturbative effects such as
renormalons or confinement are important in almost all processes. A more
useful alternative would be a formulation where both confinement and partons
are incorporated at ``tree" level, as defined by partons or hadrons appearing
only as poles, with small corrections from loops, without the need for further
contributions that are both poorly defined and almost impossible to
calculate. There is some experimental evidence to indicate that such an
approach is possible \cite{vg,dok,hoyer}. In this paper we propose such a model, and give
some preliminary comparison to the real world.

As we consider only the tree
contribution, we compare mostly to reggeons, since for the pomeron
(``glueballs") cuts are difficult to disentangle from poles for $t<0$, while no glueballs have been unambiguously identified to allow identification of trajectories for $t>0$.

%_________________________________________________________________________________
\mysection{Cuts and sisters}

In this section we give some background on the cut and sister trajectory
approaches to high-energy behavior to make the further discussion more
tangible. In both the approaches, the main idea is that a resummation of
string loops will modify the effective leading trajectory: In the cut approach
one looks at the trajectories of the
cuts produced by exchanges of multiple poles; similarly, in the sister
trajectory approach, the sister trajectories appear only in
higher-and-higher-point amplitudes, so a summation over these trajectories can
be applied only by a summation over all loops.

In the case of cuts, the general rule for linear trajectories (poles or cuts), written as
\begin{equation}\label{alpha}
\alpha (t) = \frac{\alpha'}{a}t - b + 1
\end{equation}
is that the trajectory $\alpha$ resulting from the exchange of multiple
trajectories $\alpha_i$ satisfies
\begin{equation}\label{alpha1}
a = \sum a_i\quad,\quad b = \sum b_i
\quad.
\end{equation}
For example, if we consider the sum of $n$ identical trajectories (in units $a_i=1$), we find
\begin{equation}\label{alpha2}
\alpha_n (t) = \frac{\alpha'}{n}t - bn + 1
\quad.
\end{equation}
(We could also consider one reggeon trajectory plus $n-1$ pomeron
trajectories, with qualitatively similar results.) Sister trajectories have a similar form,
\begin{equation}\label{alpha3}
\alpha_n (t) = \frac{1}{n}\bigl(\alpha't +\alpha_0\bigr)- \frac{1}{2}\bigl(n-1\bigr)
\quad.
\end{equation}

The basic idea is that in general one trajectory will be higher than the rest,
where the value of $n$ for that trajectory will depend on the value of
$t$. Explicitly, if we sum the high-energy contributions over $n$,
\begin{equation}\label{alpha4}
\sum_n \beta_n (t)\bigl(\alpha_n(s)\bigr)^{\alpha_n(t)}
\quad,
\end{equation}
then the leading contribution can be found by a saddle-point approximation on
the exponent of $\alpha_n(s)$, approximating the sum as an integral.\footnote{A better
approximation would include the effect of $\beta_n$ and the $n$-dependence of
$\alpha_n(s)$, as we'll see in the following section.}
The result of treating $n$ as continuous is that a differentiable curve is obtained
for this effective trajectory, which is a better approximation than the piecewise
differentiable trajectory that would be obtained by simply connecting together
the pieces of whichever trajectory happens to be highest in any particular region.
For a generic
contribution that includes the cut and sister cases,
\begin{equation}\label{alpha5}
\alpha_n (t) = \frac{\alpha'}{n}\bigl(t -t_0\bigr)-bn+J_0
\quad,
\end{equation}
where $b\ge 0$, we find the maximum (for $t<t_0$) from
\begin{equation}\label{alpha6}
0 = \frac{\partial}{\partial n}\alpha_n (t) = -\frac{\alpha'(t-t_0)}{n^2}-b
    \quad\Rightarrow\quad n_0 = \sqrt{\alpha'(t_0 -t)/b}
\end{equation}
\begin{equation}\label{alpha7}
\Rightarrow\quad \tilde\alpha (t) =
\begin{cases}
\alpha_1 (t) = \alpha'(t-t_0)-b+J_0\quad&\text{for}\quad t\ge t_0-b \\
\alpha_{n_0}(t) = -2\sqrt{\alpha'b(t_0 -t)}+J_0\quad&\text{for}\quad t\le t_0-b
\end{cases}
\quad,
\end{equation}
where $\tilde\alpha$ is the ``top trajectory" obtained by combining the parts
of the trajectories from each $n$ where it is greater than the others.  This
modifies the effective behavior of the amplitude, but not enough to mimic
parton behavior in the hard limit. The only exception is the case $b=0$: This
is irrelevant to the sister case, while in the cut case it relates to the
pomeron, whose intercept is near 1, with trajectories converging at
$t=0$. That case is too extreme, since it would eliminate Regge behavior
altogether (flat trajectory for all $t<0$).

%__________________________________________________________________________
\mysection{New models}

Our models will be based on several assumptions:
\newline (i) {\it Amplitudes are sums of ``standard" (primary) string amplitudes.} In this
paper we examine only the 4-point amplitudes, so this means Beta functions, or
more general ratios of products of Gamma functions, whose arguments are linear
trajectories. This guarantees duality.
\newline (ii) {\it All amplitudes have poles that are a subset of those of the
``first'' amplitude.} This prevents cuts in this Born approximation (unlike
continuously running tension). The trajectories are then quantized, so we
parameterize them by a positive integer ``$n$", with ``first" meaning $n=1$.
For simplicity we assume no degeneracy, so this one parameter is sufficient to
identify a trajectory. (In principle, degeneracy might be hidden in the
normalization of the weights.)
\newline (iii) {\it The trajectories converge toward a flat trajectory.} This allows
parton behavior \cite{cim}, since reggeons with small slopes resemble ordinary particles
for a large range of energy. The natural ordering is for the trajectories'
slopes to decrease with increasing $n$.  For simplicity we assume the slopes
are non-degenerate. Thus the slopes approach zero in the limit as $n$ goes to
infinity. The intercepts also converge (unlike methods using cuts or the usual
sister trajectories), so the ``top trajectory" approaches a constant at
$t=-\infty$. The use of an infinite number of trajectories is also a
simplifying assumption, since it allows the small-slope contribution to be
approximated by an integral: Integrating over slope (tension) produces
approximate conformal invariance at large transverse momenta.
\newline (iv) {\it The weights are also $n$-dependent, in a way consistent
with quark counting rules in the hard scattering limit where they are relevant.}

To preserve the integer (in units $\alpha'=1$) spacing of the poles, we
require the masses of the states on the leading (linear) trajectory for each
$n$ satisfy
\begin{equation}\label{traj}
\alpha'M^2 + \alpha_0 = a_nJ + b_n\quad\text{for mass}\,\,\,M\,\,\, \text{and spin}\,\,\, J
\quad,
\end{equation}
where $\alpha_0$ is an $n$-independent constant (determined by the trajectory
for $n=1$), and $a_n$ and $b_n$ are $n$-dependent integers. Since these states
appear at integer $J=\alpha (M^2)$, we have for the trajectories
\begin{equation}\label{traj1}
\alpha_n (t) = \frac{1}{a_n}\bigl(\alpha't + \alpha_0-b_n\bigr)
\quad,
\end{equation}
where $a_n$ increases with increasing $n$.
(We can always normalize $\alpha_0$ so $b_1=0$; in most cases we also have $a_1=1$, so $\alpha_0$ is
the intercept for $\alpha_1(t)$.) Since the trajectories converge
\begin{equation}\label{a/b}
\lim_{n\to\infty}\frac{b_n}{a_n}=\text{const}
\end{equation}
power-law behavior in the hard scattering limit can be obtained by choosing
the relative normalization of the weights for the amplitude $\hat A$ so that
\begin{equation}\label{amp}
\hat A=\sum_{n=1}^\infty w_n\, A(n)
\quad,
\end{equation}
where $w_n$ is a weight and $A(n)$ differ only by the fact that they depend on
$\alpha_n$. One convenient choice is to take $w_n$ in the form\footnote{Note that any function
whose large-$n$ asymptotics has such a form provides the parton behavior in the hard scattering limit.}
\begin{equation}\label{weight}
w_n=\frac{c}{n}\,a_n^{-c}
\quad,
\end{equation}
where $c$ is some parameter, since this form is less sensitive to choice of
$a_n$:  When we approximate the sum as an integral for large $n$, if we choose
$a_n$ to go as a power of $n$ in that limit, then different choices of that
power will have little affect on the integral, as it depends on $n$ only
through $a_n$, $b_n$, and the ``measure" $dn/n$.  From a hard scattering analysis, $c$ will turn
out to be integer, half the total number of quarks.  ($c=4$ for the 4-meson amplitude on which we focus.)

%______________________________________________________________________________
\mysection{Backgrounds}

We have not given the physical interpretation of the integer parameter
$n$. One possibility is first-quantization: It might be the zero-mode
(perhaps the only mode) of a fifth dimension, whose momentum is
compact.\footnote{A topological quantity such as worldsheet instanton number
  or Euler number does not seem possible, since those are worldsheet-integrated quantities.}
If so, it would be interesting to see a relation to the model of \cite{nah}
where the extra dimension is also latticized.
Alternatively, it might be a consequence of latticization of the worldsheet itself:
Random lattice quantization, before taking the worldsheet continuum limit, can lead to quantized values of the slope \cite{bgs}.

Another possibility is second-quantization:  The slope, intercept, and string coupling
are quantized, suggesting something along the line of the quantization of
the gravitational constant found in another context by Bagger and Witten
\cite{edd}.  In this interpretation summation over the values of these
couplings, like a sum over instantons, would initially be considered
nonperturbative; the result of this simple resummation would then be
treated as the tree approximation of a new perturbation expansion. The
definitions of ``nonperturbative" and ``tree" are mere semantics; what
matters is that our definition of ``tree" gives a simple amplitude that
one can apply explicitly.

Each of the integers $a_n$, $b_n$, and $c$ would then be associated with the quantization of the ``vacuum"
value of a closed-string field:  The slope (associated with the integers $a_n$) with the (four-dimensional) ``graviton",
the intercept (associated with the ratio $b_n/a_n$) perhaps with the ``tachyon", and $c$, which is required by dimensional analysis in terms of the number of quarks, with some other scalar, like the ``dilaton" (which by definition is related to dilatations and thus engineering dimension) or a higher-dimensional component of the metric.  (Of course, for the
pomeron, or closed hadronic string, all these states are now massive; we simply use the names associated with
these fields in conventional string theory.)

Since $a_n$ and $b_n$ are functions of $n$, the first step would be to find  background fields,
representing a ``ground state" solution of some field equations about which string perturbation is performed,
that are functions of a fifth dimension $r$ such that the fields take integer values when the ``warp factor"
$a(r)$ does:  For example, it appears in the spacetime metric, and thus the string Lagrangian, as
\begin{equation}\label{lag}
ds^2 = -a(r) dx^2 + ...
\quad\Rightarrow\quad
L = \frac{a(r)}{\alpha'_{10}}(\partial x)^2 + \cdots
\end{equation}
where $\alpha'_{10}$ is the usual slope of the 10D string.
(We discard terms for other
coordinates by assuming that they are $x$-independent.)
The amplitude is then defined by
\begin{equation}\label{back2}
\hat A_4 (\alpha') = \int_{r_0}^\infty \frac{dr}{r} \,ca(r)^{-c}\,
A_4 \left(\frac{\alpha'_{10}}{a(r)} \right) \quad\quad , \quad\quad
\alpha' =\frac{\alpha'_{10}}{a(r_0)} \quad .
\end{equation}
For example, the case of AdS${}_5$\cite{ps} in Eq.\eqref{am} has
$a(r)=r^2/R^2$, where $R$ is the radius of AdS${}_5$.  As a
result, all parameters in the 10D formulation ($\alpha'_{10}$,
$r_0$, $R$, 10D string coupling) are replaced by just the 4D
string coupling and slope $\alpha'$.

The next step would be to replace this integral with the sum in Eq.\eqref{amp} by
performing the quantization
\begin{equation}\label{quant}
 \frac{a[r(n)]}{a(r_0)} = \frac{a_n}{a_1}
 \quad,
 \end{equation}
where $r=r_0$ corresponds to $n=1$.  However, the dependence of the fields on $r$ should be consistent with the
relations between $a_n$ and $b_n$ that we have already found.  Effectively, this is the same as looking at the
large-$n$ limit of our models, and treating the primary amplitudes as functions of continuous $n$ (i.e., $r$)
in this limit.  This limit will also be important below in analyzing the high-energy limits of amplitudes.

This restriction eliminates certain types of backgrounds found in many supergravity solutions: For instance,
in the D3 brane solution there is the nonconformal geometry
\begin{equation}\label{back1}
a(r)=\sqrt[k']{c_0+c_k\Bigl(\frac{r}{R}\Bigr)^k}
\quad,
\end{equation}
with some integers $k$ and $k'$ ($\ne 1$), which will not lead to integer $a_n$.
Also, we have not included the
harmonic functions usually occurring in supergravity solutions.
A true string derivation of our models will require understanding the origin of both background and quantization.

We have not taken into account the affects of (broken) supersymmetry:  In particular, we have not considered Ramond-Ramond background fields, which would require a Green-Schwarz formulation.

%______________________________________________________________________________
\mysection{Regge limit}

\subsection{Lowest order approximation}

As a simple example, we apply the analysis of section 2 to a trajectory
\begin{equation}\label{an}
\alpha_n(t)=\frac{\alpha'}{a_n}(t-t_0)+J_0
\quad,
\end{equation}
for some integer $J_0$.
The leading intercept takes the arbitrary value $J_0-\alpha' t_0/a_1$, but for large $n$ the intercepts converge to $J_0$.\footnote{This includes as a special case the model of \cite{ps}, where $t_0=0$, and $J_0=2$ is the usual closed string intercept, for each trajectory.}  Since by assumption $a_n$ being positive increases
indefinitely as $n$ increases indefinitely, we easily obtain
\begin{equation}\label{an1}
\tilde\alpha(t)=
\begin{cases}
\alpha_1(t)=\frac{\alpha'}{a_1}(t-t_0)+J_0\quad&\text{for}\quad t\ge t_0\,\,,\\
\alpha_\infty(t)=J_0\quad&\text{for}\quad t\le t_0\,\,.
\end{cases}
\end{equation}
We can thus arbitrary fit the point $(t_0,J_0)$ where $\tilde\alpha$ goes
from flat to slope $1$.  A particularly simple case is $a_n=n$.

By a slight generalization of the case $a_n=n^2$, the asymptotic intercept can be generalized to half-integer, and the top trajectory can be made more smooth:
\begin{equation}\label{hyp}
\alpha_n (t)=\frac{\alpha'}{n^2}\bigl(t - t_0\bigr)+\frac{1}{n}\Bigl(J_0
-\frac{1}{2}J_1\Bigr)+\frac{1}{2}J_1
\quad,
\end{equation}
where $J_0$ and $J_1$ are some integers obeying $J_1\le 2J_0$. The top trajectory is then given by
\begin{equation}\label{hyp2}
\tilde\alpha(t)=
\begin{cases}
\alpha_1 (t) = \alpha'\bigl(t-t_0\bigr)+J_0\quad&\text{for}\quad t\ge
t_0-
\frac{1}{2}(J_0 -\frac{1}{2}J_1)\,\,, \\
\alpha_{n_0} (t) = (J_0 -\frac{1}{2}J_1)^2/4\alpha'(t_0-t)+
\frac{1}{2}J_1\quad&\text{for}\quad t\le
t_0-\frac{1}{2}
\bigl(J_0 -\frac{1}{2}J_1\bigr)\,\,,
\end{cases}
\end{equation}
which replaces the flat part with a hyperbola. The extra parameter over the
previous case allows for choice of the sharpness of the hyperbola, which
allows a smoother transition to flatness. (The previous top trajectory is
obtained for $J_1=2J_0$.)

As a further generalization, consider
\begin{equation}\label{an2}
\alpha_n(t)=\frac{1}{n^k}\bigl(\alpha't+\alpha_0-P_k\bigr)
\quad.
\end{equation}
of which the previous example is the special case $k=2$.
Its top trajectory is given by
\begin{equation}\label{an3}
\tilde\alpha(t)=
\begin{cases}
\alpha_1(t)=\alpha't+\alpha_0-P_k(1)
\quad&\text{for}\quad n_0\le 1\,\,,\\
\alpha_{n_0}(t)=-P'_k(n_0)/kn_0^{k-1}
\quad&\text{for}\quad n_0\ge 1\,\,,
\end{cases}
\end{equation}
where $P_k$ is a polynomial of degree $k$ with positive coefficients. $n_0$ is
a solution of the equation $\alpha't+\alpha_0=P_k-\frac{n}{k}P'_k$. Since the
right hand side of this equation increases with increasing $n$ for $n>0$, the
solution exists if $P_k(0)<\alpha't+\alpha_0$. Note that other exactly
solvable examples are those of $k=3$ and $k=4$.

The story becomes more and more involved when effects of $\beta_n$ and
$\alpha_n(s)$ are taken into account. The novelty is that $n_0$ depends on
$s$ in a way that restricts the Regge behavior to special kinematical regions. On
the technical side, a difficulty is related to the problem of solving the
equation for the top trajectory. The example of \cite{ps} includes
simple power functions for $w_n$ and $a_n$.

\subsection{Continuous limit}

The approximation of simply determining the top trajectory works well for values of $t$ where
$\tilde\alpha(t)=\alpha_1(t)$.  In general this means for positive $t$, where the trajectories are
fit to the spectrum, but can be extended some distance to negative $t$ (e.g., by choice of the
parameter $t_0$ in the above examples).

Since experiments have not yet determined dependence on $t$ for a large range of negative
values (in comparison with that for positive $t$), that may be sufficient. However, if we anticipate
restrictions on possible models from criteria we have not yet analyzed (higher-point functions, loops,
etc.), it will be useful to generalize by considering corrections to Regge behavior from $n$-dependence
of the couplings that weight the primary amplitudes.  This was found to be the case in \cite{ps}, where
the flat part of the top trajectory found in the first example above was found to have effective
nonvanishing slope for a region consistent with experiment.

For the first model of the previous subsection, using the couplings of \eqref{weight} with Veneziano
amplitudes as the primary amplitudes,\footnote{We can also include kinematic factors, which we
assume are $n$-independent.}
\begin{equation}\label{aml}
\hat A = \sum_{n=1}^\infty \frac{c}{n}\,a_n^{-c}
\int_0^1 du\ u^{S/a_n +k}(1-u)^{T/a_n +k}
\quad.
\end{equation}
Here $S = -\alpha'(s-t_0),\,\,T = -\alpha'(t-t_0),\,\,k = J_0 -1$. We first replace the sum with an
integral, and make the change of variables
\begin{equation}\label{newv}
v = \frac{1}{a_n}\quad.
\end{equation}
If we assume $a_n$ goes as a power of $n$ (which we can normalize as $a_1=1$), then, dropping an
overall constant,
\begin{equation}\label{aml1}
 \hat A \approx \int_0^1 dv\, cv^{c-1} \int_0^1 du\ u^{Sv +k}(1-u)^{Tv +k}
 \quad.
 \end{equation}
Since we are looking for Regge behavior with nonvanishing slope, we will assume that for
large $s$ the integral over $v$ is dominated by $v\approx 1$ (small $n$), and see under what
conditions this assumption is justified.\footnote{See also \cite{oa}.} We therefore rearrange this
integral as
\begin{equation}\label{aml2}
\hat A \approx \int_0^1 dv\, cv^{c-1} \int_0^1 du\ u^{S +k}(1-u)^{T +k}
\text{e}^{-(1-v)[S\ ln\ u +T\ ln(1-u)]}
\quad.
\end{equation}
Since $S$ and $T$ are linear in $\alpha'$, the exponential can conveniently be rewritten in terms
of derivatives with respect to $\alpha'$, allowing the $u$ and $v$ integrals to be separated. The $u$
integral can then be identified as the Veneziano amplitude for the first primary amplitude,
yielding the expression
\begin{equation}\label{aml3}
\hat A \approx f\Bigl(\alpha'\frac{\partial}{\partial\alpha'}\Bigr)
B(-\alpha_1(s),-\alpha_1(t))
\quad,
\end{equation}
with
\begin{equation}\label{aml4}
f(x)=c\int_0^1 dv\, v^{c-1} \text{e}^{-(1-v)x} =
\sum_{n=0}^\infty \frac{c!}{(c+n)!}\, (-x)^n
= 1 -\frac{1}{c+1}\,x +O(x^2)
\quad.
\end{equation}
After taking the Regge limit, we have
\begin{equation}\label{aml5}
\hat A \approx f\Bigl(\alpha'\frac{\partial}{\partial\alpha'}\Bigr)
\Gamma(-\alpha_1(t))(-\alpha_1(s))^{\alpha_1(t)}
\quad.
\end{equation}
The modification to the $s$ dependence comes from the derivatives acting on the $\alpha'$ in the
exponent of $s$, so the first two terms in the expansion of $f$ yield
\begin{equation}\label{aml6}
\hat A \sim \biggl(1+\frac{\alpha'(t_0-t)}{c+1}\ln(\alpha's)\biggr)
(\alpha's)^{\alpha_1(t)}
\quad.
\end{equation}
We thus see that the range of validity of $\alpha_1$ as the effective Regge trajectory is extended from
the region $t\ge t_0$ found in the previous subsection to the additional region (in $t<t_0$)
\begin{equation}\label{aml7}
 t_0 - t \ll \frac{c+1}{\alpha'\ln(\alpha's)}
 \quad.
 \end{equation}

%______________________________________________________________________________
\mysection{Hard scattering limit}

We begin by writing a tree Neveu-Schwarz amplitude for massless vectors
\begin{equation}\label{4v}
A^{(0)}_4(\alpha')=\bigl(\alpha'\bigr)^2K\,
\frac{\Gamma(-\alpha's)\,\Gamma(-\alpha't)}{\Gamma(1-\alpha's-\alpha't)}
\quad,
\end{equation}
with the usual kinematical factor $K$ (see, e.g., \cite{jhs}). In general, a modified
amplitude will have a subset of poles of the primary amplitude \eqref{4v}, if $\alpha'$ is
replaced with $\alpha'/a_n$ such that the function $a_n$ takes only positive
integer values.\footnote{This is not quite the same as Eq.\eqref{an}: $t_0=0$
  and $\alpha_0=1$. However, in the hard scattering limit it doesn't
  matter. We will have more to say on this subject below.} Take, for example,
a polynomial of degree $k$ with positive
integer coefficients $P_k(n)$. According to our ansatz \eqref{amp}, the
modified amplitude is then
\begin{equation}\label{4v1}
\hat A_4^{(0)}=\sum_{n=1}^\infty w_n\, A^{(0)}_4(\alpha'/P_k)
\quad.
\end{equation}
For what follows we assume that $w_n$ is a product of power functions like
$n^\delta P_k^\gamma$.

To evaluate the amplitude in the hard scattering limit, $s\rightarrow\infty$, $s/t$ fixed, we first split 
the sum into two parts and then replace the second sum with an integral as
\begin{equation}\label{4v2}
\hat A_4^{(0)}=\sum_{n=1}^{[n_c/N]} w_n A^{(0)}_4(\alpha'/P_k)+
\int^\infty_{[n_c/N]} dn\, w_n A^{(0)}_4(\alpha'/P_k)
\quad,
\end{equation}
where $n_c$ is a solution of equation $P_k(n)=\alpha' s$. For this value of $n$
the arguments of gamma functions are of order $1$, so Stirling formula is not
applicable. Note that $n_c\sim\sqrt[k]{\alpha's}$ for $\alpha's\rightarrow\infty$. $[x]$ means the
integer part of $x$. $N$ is a free parameter such that Stirling formula is
applicable for all the terms of the sum. If so, then the sum provides
exponential falloff in the hard scattering limit. To see that the integral
provides the desired power law, it is enough to rescale $n$ as
$n\rightarrow\sqrt[k]{\alpha's}\,n$. Indeed, in the lower integration limit a
factor $\sqrt[-k]{\alpha's}$ cancels out the leading one from $n_c$. So, it
behaves as $const+O(1/\sqrt[k]{\alpha's})$. As to the integrand, we have
\begin{equation}\label{int}
\alpha_n(s)=\frac{1}{c_kn^k}
\Bigl(1-\frac{c_{k-1}}{c_kn\sqrt[k]{\alpha's}}+\dots\Bigr)
\quad.
\end{equation}
Finally, the amplitude behaves as
\begin{equation}\label{4v3}
\hat A_4^{(0)}\sim \bigl(\alpha's\bigr)^{2+\gamma+(\delta+1)/k}
\Bigl(1+O\bigl(1/\sqrt[k]{\alpha's}\,\bigr)\Bigr)
\quad.
\end{equation}
We have used that $K\sim s^2$ in the hard scattering limit.\footnote{As in QCD \cite{bf}, 
this is due to scattering of vector particles.}

To compare with hard processes in QCD, we note that the corrections to the
scaling behavior correspond to sea quarks and go as $1/s$. Thus, it seems
to be reasonable taking the polynomial in the following form
$P_k(n)=c_kn^k+c_0$. The other parameters can be fixed by noting that
QCD amplitudes scale as $s^{2-\text{n}/2}$, where n is a total number of
valence quarks. We take the option $\delta=-1$ and $\gamma=-\text{n}/2$
(see \eqref{weight}). A significant difference from others is that
ours is universal for all values of $k$.

It is also worth looking at a pole structure of the amplitude
\eqref{4v1}. Using $\delta=-1$ and $\gamma=-\text{n}/2$, we find
\begin{equation}\label{poles}
\hat A_4^{(0)}=\bigl(\alpha'\bigr)^2K\,
\sum_{n=1}^\infty\sum_{m=0}^\infty\,
\frac{1}{n}P_k^{-1-\text{n}/2}\,\frac{1}{\alpha's-mP_k}\,
\,\frac{\bigl(1+\alpha't/P_k\bigr)_{m-1}}{m!}
\quad,
\end{equation}
where $(x)_n$ stands for a Pochhammer polynomial. This equation shows that the
poles are indeed a subset of those of the primary amplitude \eqref{4v} and
their distribution is a function of two integers $(n,m)$. The residue of
$\hat A_4^{(0)}$ at $\alpha's=l$ is given by
\begin{equation}\label{poles1}
\gamma(l)=\frac{\alpha'}{t}K\,\sum_{\{n\}}\,
\frac{1}{nl}\,P_k^{1-\text{n}/2}\,\text{B}^{-1}(\alpha't/P_k,l/P_k)
\quad,
\end{equation}
where $\{n\}$ is a set of integer solutions of the equation $l=m(c_kn^k+c_0)$. If
the solutions don't exist, then $\gamma\equiv 0$.

A pole at $l=0$ is special because all primary amplitudes contribute. From
this point of view it can be called infinitely degenerate, while all others as
finitely degenerate. The residue is
\begin{equation}\label{poles2}
\gamma(0)=\frac{\alpha'}{t}K\,\sum_{n=1}^\infty\,
\frac{1}{n}\,P_k^{-\text{n}/2}
\quad
\end{equation}
which is finite for positive $\text{n}$ as it should be.\footnote{Note that
in order that the sum be convergent $a_n$ must increase for large $n$ (see
Eq.\eqref{traj1}).} This pole corresponds to
a massless ground state similar to that of the primary amplitude. The first
massive state is due to a pole at $l=c_k+c_0$. Note that one can change its
mass by varying the parameters $c_k$, $c_0$ but keeping
$\alpha'$ close to the Planck length. The effect is similar to that of
\cite{ps}. This gives a hint that spacetime geometry of our models might be
warped.

So far we have made the simplest modification $\alpha'\rightarrow\alpha'/a_n$
of the first amplitude. The reason for doing so is that the slope is a
dimensionful parameter which is easy to trace. On the other hand, the intercept is
dimensionless, which makes it impossible to trace in kinematical
factors $K$.\footnote{A related reason is that $\alpha'$ may be associated with a
background metric, while it is unclear with which backgrounds may be
associated $\alpha_0$. One could think it of as a modulus corresponding to a ground
state mass.  In subcritical strings, the intercept may be related to other factors, such as the spacetime dimension, or coefficients of Liouville terms, which may in turn be related to a background ``tachyon".} To bypass the $K$'s, without
losing generality consider the bosonic Lovelace-Shapiro amplitude \cite{love,sh}
\begin{equation}\label{4s}
A_4^{(0)}(\alpha(s),\alpha(t))=\frac{\Gamma\bigl(1-\alpha(s)\bigr)
\Gamma\bigl(1-\alpha(t)\bigr)}
{\Gamma\bigl(1-\alpha(s)-\alpha(t)\bigr)}
\quad,
\end{equation}
where $\alpha(x)=\alpha_0+\alpha'x$. Formula \eqref{amp} then requires
\begin{equation}\label{4s1}
\hat A_4^{(0)}=\sum_{n=1}^\infty w_n
\,A_4^{(0)}(\alpha_n(s),\alpha_n(t))
\quad,
\end{equation}
where $\alpha_n(x)$ is given by Eq.\eqref{traj1}. This amplitude has a
subset of poles of the primary amplitudes if and only if $a_n$ and $b_n$ take
positive integer values. It seems natural to specialize to polynomials with integer
coefficients, say, $a_n=P_k(n)$ and $b_n=P_{k'}(n)$ whose degrees are $k$ and
$k'$, respectively. As to $w_n$, we take it as a product of power functions
$w_n=n^\delta P_k^\gamma P_{k'}^\lambda$.

To evaluate the amplitude in the hard scattering limit we proceed as
before. So, we first split the sum into two parts and then trade a second sum
for an integral \footnote{Note that in the center of mass frame,
  $t\approx-s\cos^2\frac{\phi}{2}$ and $u\approx -s\sin^2\frac{\phi}{2}$.}
\begin{equation}\label{4s2}
\hat A_4^{(0)}=\sum_{n=1}^{[n_c/N]} w_n A^{(0)}_4
\bigl(\alpha_n(s),\alpha_n(-s\cos^2\phi/2)\bigr)+
\int^\infty_{[n_c/N]} dn\, w_n A^{(0)}_4\bigl(\alpha_n(s),
\alpha_n(-s\cos^2\phi/2)\bigr)\bigr)
\quad,
\end{equation}
where $n_c$ is a solution of equation $P_k(n)=\alpha's$. $N$ is a free
parameter such that Stirling formula is applicable for all the terms of the sum. Thus
the sum provides exponential falloff. To evaluate the integral we rescale $n$ as
$n\rightarrow\sqrt[k]{\alpha's}\,n$. In the lower integration limit a factor
$\sqrt[-k]{\alpha's}$ cancels out the leading one of $n_c$. For the integrand, we obtain
\begin{equation}\label{alpha-s}
\alpha_n(s)=\frac{1}{c_kn^k}\biggl(1-\frac{c_{k-1}}{c_kn\sqrt[k]{\alpha's}}+\dots\biggr)
\biggl(1+\frac{\alpha_0}{\alpha's}-c_{k'}n^{k'}(\alpha's)^{\frac{k'-k}{k}}
\Bigl(1+\frac{c_{k'-1}}{c_{k'}n\sqrt[k']{\alpha's}}\dots\Bigr)\biggl)
\quad.
\end{equation}
Since $n$ is bounded from below and $\alpha's$ is large, we may treat subleading terms as
corrections. As noted above, the corrections to the scaling behavior in QCD go
as $1/s$. It follows that rational powers are not allowed. If so, then
$c_{k-1}=\dots=c_1=0$, $c_{k'-1}=\dots=c_{1'}=0$, and $k=k'$. In other words,
both polynomials look very similar: they contain only the leading and constant
terms and have the same degree. As a consequence, we recover Eq.\eqref{a/b} as
expected.

Finally, we have
\begin{equation}\label{4s3}
\hat A_4^{(0)}\sim \bigl(\alpha's\bigr)^{\gamma+\lambda+(\delta+1)/k}
\Bigl(1+O\bigl(1/\alpha's\bigr)\Bigr)
\quad.
\end{equation}
By comparison with the known results of \cite{br} we fix $\delta=-1$ and
$\gamma+\lambda=2-\text{n}/2$. Since the form of the polynomials is very
restricted it makes no difference if we take $w_n$ in the form
$n^{-1}P_k^{2-\text{n}/2}$ (see \eqref{weight}).

%___________________________________________________________________________
\mysection{Hadronic mass relations}

A concrete, spectacular success of the early days of dual resonance models is
that of \cite{love,avw}. Combining the Veneziano type formulae for scattering
amplitudes with the Adler condition, they found many mass relations that
agree well with experiment. It seems natural to check whether the models of
interest allow those relations too.

We begin by discussing $\pi\pi$ scattering along the lines of
\cite{love}. Consider the amplitude \eqref{4s1}. The Adler condition requires
the amplitude to vanish when $s=t=u=m^2_\pi\approx 0$. Assuming that there is
no cancellation between different terms, we get from the denominators
\begin{equation}\label{rho}
\alpha_n(0)=\frac{1}{2}
\quad.
\end{equation}
The novelty is the $n$-dependence. For the trajectory with
$n$-independent intercept like \eqref{an} with $t_0=0$, the $n$-dependence is
in fact missing. As a result, the trajectory obeys this requirement as in the usual case, i.e., if
$\alpha_0=1/2$. It gives the intercept of the $\rho$ trajectory.
For the trajectory like \eqref{traj1}, we conclude that
\begin{equation}\label{rho1}
\alpha_0=\frac{1}{2}\,a_n+b_n
\quad.
\end{equation}
Since $\alpha_0$ does not depend on $n$, Eq.\eqref{rho1} shows that $a_n$ and
$b_n$ must be integers of opposite signs. If so, the Adler condition provides
the constraint on $b_n$. Inserting it back into Eq.\eqref{traj1} we get the
trajectory discussed before.

We should caution the reader that in principle the amplitude can take the form
$\hat A^{(0)}_4=(s+ct)f(s,t)$ with $f(0,0)\not=0$. In this case the above
derivation of the intercept fails.

It is straightforward to extend the above analysis to the case when all
particles but one to be arbitrary hadrons $\pi +A\rightarrow B+C$
\cite{avw}. Assuming that amplitudes receive
contributions from only one family of trajectories in each channel, the
amplitude to be considered is given by Eq.\eqref{4s1} with $\alpha_n(s)$ and
$\alpha_n(t)$ replaced by $\alpha_n^X(s)$ and $\alpha_n^Y(s)$. Here $X$ and
$Y$ mean the corresponding families. The rest of the analysis goes along the
lines of \cite{avw}. Thus, we get
\begin{equation}\label{x}
a_n^A=a_n^X
\quad
\end{equation}
and
\begin{equation}\label{a}
\alpha_n^X(0)-\alpha_n^A(0)=\frac{1}{2}N_{AA}
\quad,
\end{equation}
with some integer $N_{AA}$. For the trajectories \eqref{an} with $t_0=0$,
Eqs.\eqref{x}-\eqref{a} show that the two trajectories must have the same slopes,
and intercepts which differ by a half-odd integer.\footnote{Note that in this
  case $N_{AA}$ is always an odd integer as in \cite{avw}.} As a consequence,
all the mass relations of \cite{avw} hold. For the trajectories \eqref{traj1},
Eq.\eqref{a} provides the constraint on the $b_n^I$'s. One possibility to
resolve it is to take $b_n^I$ in the form $b^I_n=\alpha_0-\tilde\alpha_0
a_n^I$ that immediately leads to the trajectories \eqref{an} with
$\alpha_0\rightarrow\tilde\alpha_0$. Unfortunately, we do not know all
the solutions of the constraint, so we can not answer whether all
the trajectories reduce to those of \eqref{an}.

%______________________________________________________________________
\mysection{Further Issues}

We begin with a special class of the trajectories \eqref{traj1}. It is given by 
\begin{equation}\label{par}
a_n=n
\quad,\quad
b_n=B\bigl(n-1\bigr)
\quad,\quad
\alpha_0=B
\quad,
\end{equation}
where $B$ is an integer. To make one of the possible physical interpretations of this class  somewhat clear, let us note 
that the effective tension of the  $n^{th}$ term in the series \eqref{amp} 
\begin{equation*}
T_n=Tn
\quad,\quad 
T=1/2\pi\alpha'
\end{equation*}
is nothing else but the tension of $n$ fundamental strings. If so, one can think of the series as an expansion in 
fundamental strings. After this is understood, it immediately comes to mind to consider more complicated bound 
states.  As is usual \cite{witt}, this can be done by introducing D-strings. 

In the presence of bound states $(n,m)$ ($n$ F-strings and $m$ D-strings) it seems natural to modify the 
expression \eqref{amp} as \footnote{Interestingly enough, scattering 
amplitudes of $SL(2,Z)$-covariant superstrings as suggested in \cite{cs} are given by \eqref{am1} with $w_{nm}=const$.}
\begin{equation}\label{am1}
\hat A=\sideset{}{'}\sum_{n=0,\; m=0}^\infty w_{nm}\, A(n,m)
\quad,
\end{equation} 
where the effective tension of the  $(n,m)^{th}$ term is now
\begin{equation*}
T_{nm}=T\sqrt{n^2+\frac{m^2}{g^2}}
\quad.
\end{equation*}
$g$ stands for the string coupling. There is, however, a subtle point here: according to section 3 $a_{nm}$ and $b_{nm}$ must be integers. 
A possible way to avoid this difficulty is to take the original $a_n$ and $b_n$ as even-degree polynomials and restrict $g$ to rational values. 
Since  $m^2/g^2$ must be integer, it will restrict possible values of $m$ in the sum \eqref{am1}. We will not  drill deeper into details 
leaving them for future study. 

A final remark: one surprise of $SL(2,Z)$-covariant superstrings is that the theory is in fact 12 dimensional \cite{town}. 
It lives  in a flat space with a diagonal metric taking values $\pm 1$. As known, one may think of the model \eqref{par} as 
a zero-mode approximation to string theory whose spacetime metric is warped. For example, it is given by   
\begin{equation}\label{metric}
ds^2=f(r)dx^2+dr^2+ds^2_X
\quad,
\end{equation}
where $X$ is a five dimensional compact space. It seems natural to suggest that for the models \eqref{am1} 
the corresponding metric is given by  
\begin{equation}\label{metric1}
ds^2=f(r,\bar r)dx^2+dr^2+d\bar r^2+  ds^2_{X'}
\quad,
\end{equation}
where $X'$ is now a six dimensional compact space. Note that the novelty is warping.
  
%_______________________________________________________________________
\mysection{Conclusions}

One question we have not addressed is the usual constraints at the
string-loop level on the (critical) spacetime dimension and form of the
trajectories (e.g., intercept). There might also be constraints already
at the tree level, as we have not yet examined the higher-point amplitudes.

The asymptotic flatness of the top trajectories for large negative argument
suggests a possible physical interpretation of the intercept:
If these trajectories turn flat at $t=0$ (as in AdS/CFT inspired models
for the pomeron), then the intercept is related to the effective spin at
$t=-\infty$.  If the ``state" corresponding to $t=-\infty$ is identified
with a jet, and this effective spin with the parton carrying almost all
the energy, then we expect intercept 1/2 for the reggeon (spin of that quark)
and intercept 1 for the pomeron (spin of that gluon), in qualitative agreement with
experiment.  (For the reggeon case corrections can be attributed to quark masses; for the pomeron
case there can be significant corrections due to cuts.)  In this picture it is a jet, the
experimental signature of the parton, that is treated as ``fundamental" rather than the
corresponding parton itself:  The jet is just a string in a certain off-shell kinematic limit.

These models might also be used for fundamental strings, including gravity.
The existence of parton behavior at high energies indicates the graviton would
be a bound state in a way similar to hadrons in QCD, so that gravity would
disappear at short distances once the plasma phase is reached.

%__________________       Acknowledgments   ________________________
\vspace{.5cm}
\noindent
{\bf Acknowledgments}

\vspace{.25cm}
O.A. would like to thank S. Brodsky, H. Dorn, G. de Teramond, and
A.A. Tseytlin for useful communications and conversations. W.S. thanks
M. Islam, J. Smith, and G. Sterman for discussions. The work of O.A. was supported in
part by DFG under Grant No. DO 447/3-1 and the European Commission RTN
Programme HPRN-CT-2000-00131. The work of W.S. was supported in part by the
National Science Foundation Grant No. PHY-0354776.

%__________________                      R E F S                    ______________________

%__________________________________________________________

\end{document}